\documentclass[twocolumn]{svjour3}
\usepackage[utf8]{inputenc}
\usepackage{graphicx}
\usepackage{graphics}
\usepackage{amsmath}
\usepackage{soul}
\usepackage[dvipsnames]{xcolor}
\usepackage[english]{babel}
\usepackage{bbold}
\usepackage{cite}
\usepackage{amssymb}
\usepackage[T1]{fontenc}
\usepackage{hyperref}
\usepackage{subfigure}

\newcommand{\p}{\partial}
\newcommand{\A}{{\tiny \mathrm{A}}}
\newcommand{\B}{{\tiny \mathrm{B}}}
\newcommand{\e}{\varepsilon}
\newcommand{\ev}[1]{\langle  #1 \rangle }

\begin{document}

\title{Progress on 3+1D Glasma simulations}
\author{Andreas Ipp 
\and David I. M\"uller
}                  

\institute{Institute for Theoretical Physics, TU Wien \\
Wiedner Hauptstr. 8-10, A-1040 Vienna, Austria \\
\email{ipp@hep.itp.tuwien.ac.at} \\
\email{dmueller@hep.itp.tuwien.ac.at}}

\date{Received: date / Revised version: date}

\maketitle

\abstract{
We review our progress on 3+1D Glasma simulations to describe the earliest stages of heavy-ion collisions.
In our simulations we include nuclei with finite longitudinal extent and describe the collision process as well as the evolution of the strongly interacting gluonic fields in the laboratory frame in 3+1 dimensions using the colored particle-in-cell method.
This allows us to compute the 3+1 dimensional Glasma energy-momentum tensor, whose rapidity dependence can be compared to experimental pion multiplicity data from RHIC.
An improved scheme cures the numerical Cherenkov instability and paves the way for simulations at higher energies used at LHC.}

\section{Introduction}
\label{intro}
QCD matter under extreme temperatures and densities in the form of the quark-gluon plasma is experimentally accessible in relativistic heavy-ion collisions.
The possibility to conduct these experiments for a wide range of collision energies and baryon chemical potentials allows for the successive exploration of the QCD phase diagram.
The highest collision energies have been achieved at LHC and RHIC, and lower collision energies are being explored in the Beam Energy Scan programs of RHIC \cite{Tlusty:2018rif} and upcoming programs at GSI FAIR \cite{Senger:2020pzs} and JINR NICA \cite{Geraksiev:2019fon}.
The matter created in such collisions is initially very far from an ideal thermodynamic equilibrium.
With the experimental progress also an improved theoretical understanding of the collision process from first principles is desirable.

The Color Glass Condensate (CGC) framework \cite{Gelis:2010nm,Gelis:2012ri,Gelis:2015gza} provides such a theoretical basis for describing nuclear matter at ultrarelativistic energies.
It is a classical effective field theory where hard partons within the nuclei act as sources for soft gluonic fields.
In the simplest version, describing very large nuclei, the distribution of the color charges is given by the McLerran-Venugopalan (MV) model \cite{McLerran:1993ka,McLerran:1993ni}.
The pre-equilibrium stage that is created right after the collision is characterized by longitudinal color flux tubes and has been termed the Glasma \cite{Lappi:2006fp}.
In more sophisticated models like the IP-Glasma, the color charge distribution is based on fits to deep-inelastic-scattering data \cite{Schenke:2012hg,Schenke:2012wb}.
In combination with a subsequent hydrodynamical evolution, the IP-Glasma model is able to correctly reproduce many observables, including azimuthal anisotropies or event-by-event multiplicity distributions \cite{Gale:2012rq,Snellings:2011sz}.
Nevertheless, the underlying Glasma evolution is commonly based on a boost-invariant formulation \cite{Kovner:1995ja,Kovner:1995ts,Krasnitz:1998ns,Lappi:2011ju} where incoming nuclei are assumed to be Lorentz-contracted to infinitesimally thin discs.
This assumption is justified for observables close to mid-rapidity at very high energies, but is a severe conceptual limitation when studying rapidity-dependent quantities or collisions at lower energies.

It is possible to break boost invariance by introducing
fluctuations on top of boost invariant background fields \cite{Gelis:2013rba,Fukushima:2011nq,Berges:2012cj}.
Boost invariance is also broken by the JIMWLK evolution \cite{JalilianMarian:1996xn,Iancu:2000hn,Mueller:2001uk,Ferreiro:2001qy}. Apart from recent attempts \cite{McDonald:2018wql, McDonald:2020oyf}, Glasma simulations using JIMWLK-based initial conditions still have to be performed in an effectively boost invariant manner \cite{Schenke:2016ksl}. The generated rapidity dependence of the Glasma can reproduce observables like gluon \cite{Schenke:2016ksl} and charged hadron multiplicities \cite{McDonald:2020oyf}.
However, deviations from boost invariance may already arise at the classical level if one considers nuclei with finite extent in the beam direction  \cite{Ozonder:2013moa}.
A three-dimensional formulation has been introduced by using an extended source that is not quite aligned with the light cone \cite{Lam:1999wu, Lam:2000nz, Ozonder:2012vw}, however in a way that violates the covariant conservation condition at order $g^2$.
An alternative approach to include such finite extent corrections has been developed for proton-nucleus collisions \cite{Altinoluk:2014oxa,Altinoluk:2015gia,Altinoluk:2015xuy} which however is difficult to generalize to nucleus-nucleus collisions.

In the following we review our progress towards an alternative approach of a 3+1D simulation for heavy-ion collisions including finite longitudinal extent of the nuclei \cite{Gelfand:2016yho,Ipp:2017lho,Ipp:2017uxo,Ipp:2018hai,Muller:2019bwd}.
The loss of boost invariance requires us to keep track of the hard color sources throughout the subsequent evolution after the collision.
This is achieved using the colored particle-in-cell method (CPIC), which has been originally developed to study aspects of the evolution of the quark-gluon plasma \cite{Hu:1996sf,Moore:1997sn,Dumitru:2005hj,Dumitru:2006pz,Schenke:2008gg}. We apply this technique to study the collision process itself within the CGC framework.
The simulation is performed in the laboratory frame and follows the nuclei throughout the collision process.
Using this approach, we demonstrate that already a classical leading-order CGC simulation can give rise to a rapidity dependency consistent with experimental findings.
Recent algorithmic developments will allow us to scrutinize our findings at even higher energies.

\begin{figure}
\resizebox{0.5\textwidth}{!}{
  \includegraphics{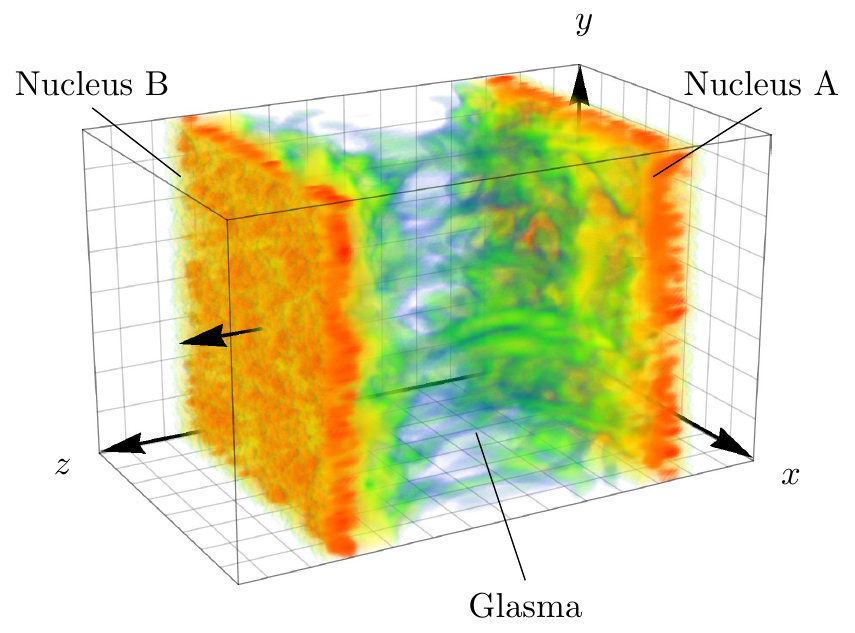}
}
\caption{
Three-dimensional simulation of the collision of two nuclei. 
The distribution of the energy density between the two nuclei ``A'' and ``B'' right after the collision reveals the flux tube structure of the Glasma that develops between them.
The simulation only covers a small part of the full collision in the transverse plane spanned by $x$ and $y$. Adapted from \cite{Ipp:2017lho}.
\label{fig:fig1}}
\end{figure}

\section{The 2+1D Glasma}

From the viewpoint of the laboratory frame, a high energy nucleus is highly Lorentz-contracted along the beam axis and its dynamics are slowed down by time dilation. At collision energies available to RHIC and LHC, nuclei therefore appear to be almost infinitesimally thin, static discs moving at highly relativistic velocities. In CGC effective theory  \cite{Gelis:2010nm,Gelis:2012ri,Gelis:2015gza}  the partons of high nuclei are split into hard and soft degrees of freedom. 
At leading order, hard partons are described in terms of classical color currents $J^\mu$ and soft partons in terms of classical color fields $A^\mu$, whose dynamics are governed by the Yang-Mills (YM) field eqs.  For example, the classical color current associated with a nucleus moving along the negative $x^3 = z$ axis (shown as nucleus ``A'' in fig.~\ref{fig:fig1}) is given by
\begin{equation} \label{eq:JA}
J_\A^\mu(x) = \delta^\mu_- \rho_\A^a(x^+, \mathbf x_T) \mathbf{t}^a,
\end{equation}
where we have used light cone coordinates $x^\pm = (x^0 \pm x^3) / \sqrt{2}$ and transverse coordinates $\mathbf x_T = (x, y)$. The matrices $\mathbf{t}^a$ are the traceless, hermitian generators of the color gauge group SU($N_c$). The color charge density is given by $\rho_\A(x^+, \mathbf x_T)$ and is assumed to be highly peaked around $x^+ = 0$ to account for the high Lorentz contraction of the nucleus. Additionally, $\rho_\A(x^+, \mathbf x_T)$ is static in the sense that it does not explicitly depend on $x^-$, which is a consequence of time dilation.

The color current $J^\mu$ induces a classical color field $A^\mu$, which is to be determined from the non-linear YM eqs.
\begin{equation} \label{eq:ym}
D_\mu F^{\mu\nu} = \p_\mu F^{\mu\nu} + i g \left[A_\mu, F^{\mu\nu} \right] =  J^\nu,
\end{equation}
where $D_\mu$  denotes the gauge covariant derivative. The non-Abelian field strength tensor is defined as
\begin{equation}
F^{\mu\nu} = \p^\mu A^\nu - \p^\nu A^\mu + i g \left[A^\mu, A^\nu \right],
\end{equation}
where $g$ is the YM coupling constant. If $J^\mu$ is given by eq.~\eqref{eq:JA}, the YM eqs.~\eqref{eq:ym} can be solved analytically by employing covariant gauge ${\p_\mu A^\mu = 0}$. In this gauge choice, the field eqs.~simplify to
\begin{equation}
- \Delta_T A_\A^-(x) =  \rho_\A^a(x^+, \mathbf x_T) \mathbf{t}^a,
\end{equation}
where $\Delta_T$ is the two-dimensional Laplace operator in the transverse plane. The other components of $A^\mu_\A(x)$ vanish. This can be formally solved by inverting the Laplace operator:
\begin{equation}
A^\mu_\A(x^+, \mathbf x_T) = \delta^\mu_- \intop^\Lambda \frac{d^2 \mathbf k_T}{(2\pi)^2}  \frac{\rho_\A^a(x^+, \mathbf{k}_T)}{\mathbf{k}^2_T + m^2} e^{i \mathbf{k_T} \cdot \mathbf{x_T}} \mathbf{t}^a.
\end{equation}
Infrared divergences are avoided by including a small regulator $m$, which dampens the long-range behavior of the color field. The length scale $m^{-1}$ is usually identified with the confinement radius. It is also possible to regulate ultraviolet modes with a cutoff $\Lambda$.
The color field of the nucleus consists of purely transverse color-electric and -magnetic fields located near $x^+ = 0$. The color field of a high energy nucleus is therefore analogous to the Lorentz-boosted electromagnetic field of an ultrarelativistic charge.

In CGC effective theory the color currents of nuclei are stochastic fields whose distribution is described by a probability functional $W[\rho]$. Expectation values of 
observables (expressed as functionals of the color fields $\mathcal O [A_\mu]$) are computed by averaging over all possible realizations of $\rho$:
\begin{equation} \label{eq:ev}
\ev{\mathcal{O}} = \int \mathcal D \rho \,  \mathcal{O}[A_\mu [\rho]] \, W[\rho]. 
\end{equation}
A simple and popular model for $W[\rho]$ is the MV model  \cite{McLerran:1993ka,McLerran:1993ni}, which assumes the functional to be Gaussian:
\begin{equation}
W[\rho] = \frac{1}{Z} \exp{\left( \! - \! \int \! d^2 \mathbf x_T d x^+ \frac{\rho^a(x^+,\mathbf x_T) \rho^a(x^+, \mathbf x_T)}{2 g^2 \mu^2(x^+, \mathbf x_T)} \right)},
\end{equation}
where $Z$ is a normalization constant and the function $\mu^2(x^+, \mathbf x_T)$ describes the variance of the randomly distributed color charges. 

The description of nucleus ``B'' (see fig.~\ref{fig:fig1}), which moves along the positive $z$ axis is completely analogous.

\begin{figure}
\resizebox{0.45\textwidth}{!}{\includegraphics{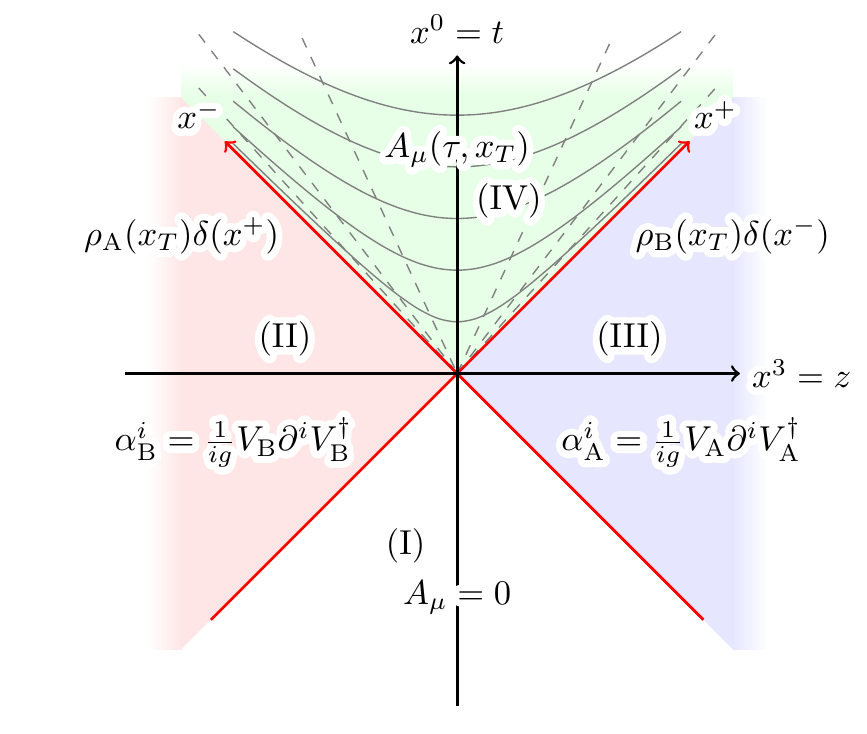}}
\caption{A Minkowksi diagram of a boost invariant heavy ion collision. The colliding nuclei move along $x^+$ and $x^-$ (red arrows) and are assumed to be infinitesimally thin along the beam axis $z$. The gray hyperbolas in region IV are contour lines of constant proper time $\tau$ and the straight dashed lines are lines of constant rapidity $\eta_s$. Adapted from \cite{Muller:2019bwd}.
\label{fig:collision}}
\end{figure}

A collision of two CGCs can be described by solving
\begin{equation} \label{eq:ym_AB}
D_\mu F^{\mu\nu}(x) = J^\nu_\A(x) + J^\nu_\B(x),
\end{equation}
where the source is given by the combined color current of the colliding nuclei. We assume nucleus ``A'' to be centered around $x^+ = 0$ and nucleus ``B'' to be centered around $x^- = 0$ such that the collision occurs at $x^+ = x^- = 0$. A Minkowski diagram of the collision scenario is shown in fig.~\ref{fig:collision}.

In the boost invariant limit, the color charge densities become effectively two-dimensional due to Lorentz contraction:
\begin{equation} \label{eq:bi_condition}
\rho_{(\A, \B)}(x) \approx \delta(x^\pm) \rho_{(\A, \B)}(\mathbf x_T).
\end{equation}
This approximation is only correct in the limit of infinite collision energy. 
In this case, the space-time shown in fig.~\ref{fig:collision} separates into four distinct regions.  Invoking causality, the solution $A^\mu(x)$ to eq.~\eqref{eq:ym_AB} in regions I - III is given by the superposition of the single nucleus solutions $A^\mu_\A(x)$ and $A^\mu_\B(x)$. The solution in the future light cone, which describes the Glasma, generally does not exist in closed form and has to be determined perturbatively or numerically. In the boost invariant limit however, the gauge field can be determined along the boundary of the light cone. Using proper time $\tau = \sqrt{2 x^+ x^-}$ and (space-time) rapidity $\eta_s = \ln \left(x^-  / x^+\right) / 2$ and employing temporal gauge $A^\tau = 0$ for $\tau \geq 0$, the color field at $\tau = 0^+$ is given by \cite{Kovner:1995ja}
\begin{align} \label{eq:glasma_ic1}
A^i(\tau = 0^+, \mathbf x_T) &= \alpha^i_\A(\mathbf x) + \alpha^i_\B(\mathbf x), \\
A^\eta(\tau = 0^+, \mathbf x_T) &= \frac{ig}{2} \left[\alpha^i_\A(\mathbf x), \alpha^i_\B(\mathbf x) \right].
\label{eq:glasma_ic2}
\end{align}
Here, the color fields $\alpha^i_{(\A, \B)}$ are the light cone (LC) gauge $A^\mp = 0$ solutions
\begin{align} \label{eq:LC_gauge1}
A^i_{(\A, \B)}(x^\pm, \mathbf x_T) &= \theta(x^\pm)  \alpha^i_{(\A, \B)}(\mathbf x_T), \\
\alpha^i_{(\A, \B)}(\mathbf x_T) &= \frac{1}{ig} V_{(\A, \B)}(\mathbf x) \p^i V^\dagger_{(\A, \B)}(\mathbf x),
\end{align}
where the lightlike Wilson lines are given by
\begin{align}
V_{(\A, \B)}^\dagger(\mathbf x_T) &= \lim_{x^\pm \rightarrow \infty} V_{(\A, \B)}^\dagger(x^\pm, \mathbf x_T), \\
V_{(\A, \B)}^\dagger(x^\pm, \mathbf x_T) &= \mathcal{P} \exp \bigg(  i g \! \! \intop_{-\infty}^{x^\pm} \! \! dx'^\pm A_{(\A, \B)}^\mp(x'^\pm, \mathbf x_T)  \bigg).
\end{align}
The initial conditions eqs.~\eqref{eq:glasma_ic1} and \eqref{eq:glasma_ic2} describe a highly anisotropic initial state consisting of purely longitudinal color-electric and -magnetic flux tubes \cite{Fries:2006pv, Lappi:2006fp}. Since these initial conditions do not depend on rapidity $\eta_s$, the Glasma and any observables remain boost invariant for $\tau > 0$.
For charge densities which are not $\delta$-shaped as in eq.~\eqref{eq:bi_condition} and instead have finite longitudinal length along $x^\pm$, there are no rigorous derivations of generalized initial conditions. In order to allow for charge densities which explicitly break boost invariance, we have to move to a fully 3+1 dimensional description of the Glasma. In particular, it is necessary to solve the YM eqs.~\eqref{eq:ym_AB} in a different way.

\section{Simulating the Glasma in 3+1D}

The motivation for relaxing eq.~\eqref{eq:bi_condition} is to go beyond the approximation of infinite collisional energy and be able to describe observables such as the energy momentum tensor of the Glasma in a rapidity dependent manner. A simple generalization is given by
\begin{equation} \label{eq:single_sheet}
\rho_{(\A, \B)}(x) \approx \lambda(x^\pm) \rho_{(\A, \B)}(\mathbf x_T),
\end{equation}
where $\lambda$ is a normalized function, which determines the longitudinal shape of the charge density. The width of $\lambda$ should be directly related to the Lorentz-contracted diameter of the nucleus. It should be noted that eq.~\eqref{eq:single_sheet} is a special case where the color structure does not depend on the longitudinal coordinate $x^\pm$ and more general color charge densities $\rho_{(\A, \B)}(x^\pm, \mathbf x_T)$ are possible as well.

A direct consequence of allowing for finite longitudinal extent is that the collision event is not just a single point in the Minkowski diagram (see fig.~\ref{fig:collision}), but an extended space-time region in which the nuclei are able to interact. The non-perturbative nature of the color fields of the nuclei and the extended interaction time generally make analytical calculations in this space-time region intractable. 
Our approach to the 3+1D Glasma is therefore to simulate not just the evolution in the future light cone, \textit{i.e.}~region IV in fig.~\ref{fig:collision}, but the whole collision \cite{Gelfand:2016yho}. This setup is formulated in the laboratory frame using $(t, z)$ coordinates and the time evolution is performed in the direction of $t$ instead of proper time $\tau$. The initial conditions of such a time evolution in $t$ are specified in the following way: at some initial time $t_0 < 0$ sufficiently far away from the collision time $t = 0$, the color charge densities of the nuclei are non-overlapping in $z$. In LC gauge, the color field of each nucleus is given by
\begin{equation} \label{eq:3d_colorfield_1}
A^{i=x,y}_{(\A, \B)}(t_0, \mathbf x) = \frac{1}{ig} V_{(\A \B)}(t_0, \mathbf x) \p^i  V^\dagger_{(\A \B)}(t_0,\mathbf  x),
\end{equation}
where $\mathbf x = (x, y, z)$ is a three-dimensional coordinate vector. Equation \eqref{eq:3d_colorfield_1} solves the classical YM eqs.~\eqref{eq:ym} with the current given by eq.~\eqref{eq:single_sheet} \cite{Gelfand:2016yho}. Since the fields vanish exponentially fast between the two nuclei, the superposition of both color fields
\begin{equation} \label{eq:3d_colorfield_2}
A^\mu(t_0, \mathbf x) = \delta^\mu_{i=x,y} \left(A^i_\A(t_0, \mathbf x) + A^i_\B(t_0, \mathbf x) \right),
\end{equation}
is a valid solution to the YM eqs.~\eqref{eq:ym_AB} at time $t_0$. Evolving this initial condition for the YM field to $t > 0$ yields a genuinely 3+1D description of the  rapidity dependent Glasma.
Equations \eqref{eq:3d_colorfield_1} and \eqref{eq:3d_colorfield_2} are compatible with the temporal gauge condition $A^0(t, \mathbf x) = 0$, $\forall t\in \mathbb{R}$, which is a convenient gauge choice for an evolution along $x^0 = t$.

One of the main differences to the standard 2+1D Glasma is that in the laboratory frame the system not only consists of the color fields but also the color currents of the nuclei. In order to solve the YM eqs.~which include color currents, we make use of the CPIC method \cite{Hu:1996sf,Moore:1997sn,Dumitru:2005hj,Dumitru:2006pz,Schenke:2008gg}. CPIC allows for consistent simulations of color charged point particles coupled to color fields on a lattice. For a comprehensive description of our numerical methods we refer to \cite{Gelfand:2016yho} and \cite{Muller:2019bwd}.

\begin{figure*}
    \centering
    \subfigure[Standard plaquette term]{\includegraphics[scale=0.92]{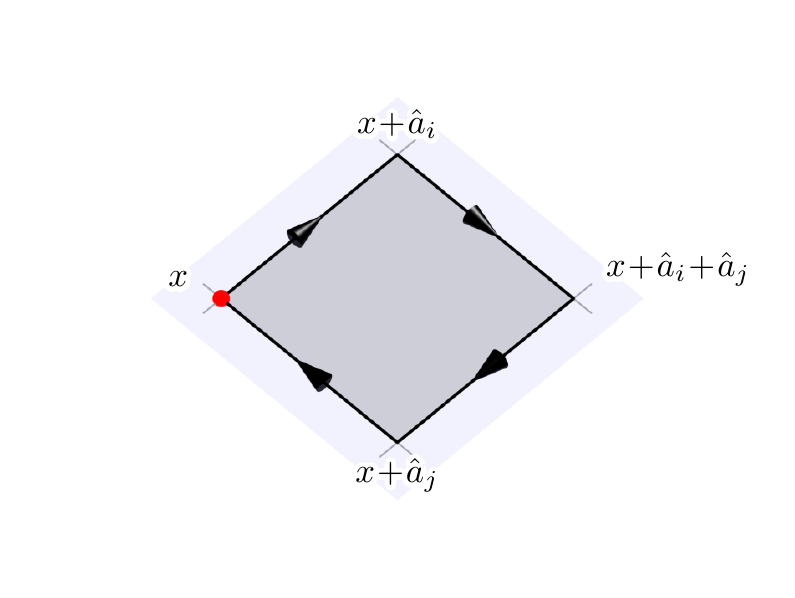}}
    \qquad
    \subfigure[Time-averaged plaquette-like term]{\includegraphics[scale=0.92]{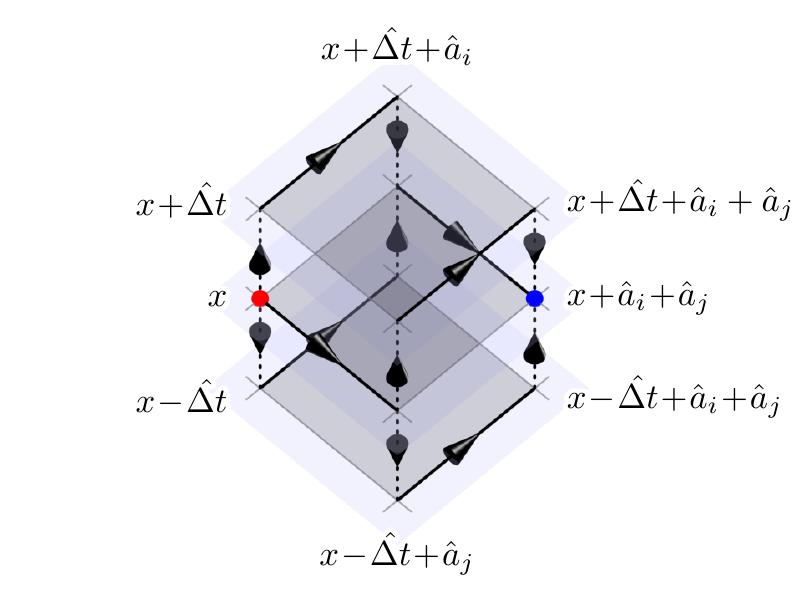}}
    \caption{Two schematic diagrams of terms used in the lattice discretization of the YM equations: the standard spatial plaquette term (a) as defined in eq.~\eqref{eq:plaquette}, which is used in the leapfrog scheme, and a time-averaged generalization of this term (b), which is used in our semi-implicit method. While (a) only contains gauge links defined at a common equal-time slice of discretized Minkowski space, the generalized term (b) includes terms from both future and past equal-time slices. Adapted from \cite{Ipp:2018hai}.}
    \label{fig:wilson}
\end{figure*}

The numerical treatment of the fields follows the common real-time lattice gauge theory approach: by discretizing Minkowski space-time as a hypercubic lattice with spacings $a^i$ and time-step $\Delta t$, the YM eqs.~can be written in the standard leapfrog scheme
\begin{align}
E_i(t+ \frac{\Delta t}{2}, \mathbf x) =& - \sum_j \frac{\Delta t}{(a^j)^2} \left[ U_{i,j}(t, \mathbf x) + U_{i,-j}(t, \mathbf x) \right]_\mathrm{ah} \nonumber \\
&+ \Delta t \, j_i(t, \mathbf x) + E_i(t - \frac{\Delta t}{2}, \mathbf x), \label{eq:lattice_1}\\
U_i(t  +  \Delta t, \mathbf x) =& \exp \bigg( i \Delta t \, E_{i}(t+\frac{\Delta t}{2}, \mathbf x) \bigg) U_i(t, \mathbf x), \label{eq:lattice_2}
\end{align}
where $U_{i}(t, \mathbf x) \simeq \exp \big( i g a^i A_i(t, \mathbf x) \big)$ are the gauge links, $E_i(t, \mathbf x)$ is the chromo-electric field on the lattice and $\left[ X \right]_\mathrm{ah}$ denotes the anti-hermitian, traceless part of a matrix $X$. The plaquette variables $U_{i,j}(t, \mathbf x)$ are defined as
\begin{equation} \label{eq:plaquette}
U_{i,j}(t, \mathbf x) = U_i(t, \mathbf x) U_j(t, \mathbf x + \hat{a}_i) U^\dagger_i(t, \mathbf x + \hat{a}_i) U^\dagger_j(t, \mathbf x). 
\end{equation}
The color currents of the nuclei $j_i(t, \mathbf x)$, which enter on the right hand side of eq.~\eqref{eq:lattice_1}, require careful treatment. The main idea of using the CPIC method to describe collisions in the CGC framework is to replace the continuous color charge distributions $\rho$ of the nuclei by a large number of auxiliary particles with time-dependent color charges $Q_k(t)$ such that the original color charge distribution is sufficiently well approximated on a lattice:
\begin{equation}
\rho(t, \mathbf x) \approx \sum_k Q_k(t) \delta^{(3)}(\mathbf x - \mathbf x_k(t)),
\end{equation}
where $k$ is the particle index. Similar to the boost invariant case, we assume these auxiliary particles to be recoil-less. Thus, the trajectories of the particles $\mathbf x_k(t)$ are fixed and not part of the dynamics of the system.
The time-dependence of the color charges $Q_k(t)$ is determined from the discretized continuity eq.
\begin{align}\label{eq:continuity}
&\frac{\rho(t  +  \frac{\Delta t}{2}, \mathbf x) - \rho(t  - \frac{\Delta t}{2}, \mathbf x)}{\Delta t} = \nonumber \\
& \quad \sum_i \frac{j_i(t, \mathbf x) - U^\dagger_{i}(t, \mathbf x  -  \hat{a}_i) j_i(t, \mathbf x -  \hat{a}_i) U_{i}(t, \mathbf x  -  \hat{a}_i)}{a^i},
\end{align}
which is the discrete analogue of gauge covariant continuity eq.
\begin{equation} 
D_\mu J^\mu(t, \mathbf x ) = 0.
\end{equation}
In our setup, the color charge $Q_k(t)$ of each particle is mapped to its nearest grid point on the spatial lattice in each time-step. Whenever the nearest grid point of a particular point charge changes from one lattice site $\mathbf x$ to a neighbouring lattice site $\mathbf y$, parallel transport is applied to the color charge accordingly:
\begin{equation}
Q_k(t  +  \frac{\Delta t}{2}) = U^\dagger_{\mathbf x \rightarrow \mathbf y}(t) Q_k(t  -   \frac{\Delta t}{2}) U_{\mathbf x \rightarrow \mathbf y}(t),
\end{equation}
where $U_{\mathbf x \rightarrow \mathbf y}(t)$ is the appropriate gauge link connecting $\mathbf x$ and $\mathbf y$. At the same time, the movement of the particle generates a color current $j_i(t, \mathbf x)$ in accordance with eq.~\eqref{eq:continuity}. In CPIC, this treatment of particles is known as the nearest-grid-point scheme \cite{Hu:1996sf}. By evolving the color charges of the particles in this manner, the discretized field eqs.~\eqref{eq:lattice_1} and \eqref{eq:lattice_2} are solved consistently in the sense that the discrete Gauss law
\begin{align} \label{eq:gauss}
&\sum_i \bigg( E_i(t  +  \frac{\Delta t}{2}, \mathbf x) - \tilde{E}_i(t  +  \frac{\Delta t}{2}, \mathbf x  -  \hat{a}_i) \bigg) = \nonumber \\
& \qquad \rho(t+\frac{\Delta t}{2}, \mathbf x) ,
\end{align}
remains satisfied throughout the simulation and gauge covariance on the lattice is retained. In eq.~\eqref{eq:gauss} the parallel transported electric field is given by
\begin{align}
&\tilde{E}_i(t  +  \frac{\Delta t}{2}, \mathbf x -  \hat{a}_i) = \nonumber \\
&\qquad U^\dagger_{i}(t, \mathbf x - \hat{a}_i) E_i(t +  \frac{\Delta t}{2}, \mathbf x  -  \hat{a}_i) U_{i}(t, \mathbf x  -  \hat{a}_i).
\end{align}

We find that numerically stable 3+1D Glasma simulations using the leapfrog scheme eqs.~\eqref{eq:lattice_1} and \eqref{eq:lattice_2} require high lattice resolution with particularly fine lattice spacing along the beam axis $z$. In practice, this can be computationally prohibitive. This problem is related to a subtle numerical instability inherent to the leapfrog scheme known as the numerical Cherenkov instability \cite{GODFREY1974504}, which also affects traditional electromagnetic plasma simulations. This unphysical instability is caused by lattice artifacts which modify the propagation of wave modes on the lattice: high frequency modes propagate at a lower phase velocity compared to low frequency modes (\textit{i.e.}~numerical dispersion). In contrast, the auxiliary particles move at the speed of light by design. This situation is reminiscent of charged particles moving through a medium in which the in-medium speed of light is lower than the particle velocity, which leads to the well-known phenomenon of Cherenkov radiation. The particular lattice discretization used in eqs.~\eqref{eq:lattice_1} and \eqref{eq:lattice_2} leads to a similar, although purely numerical generation of Cherenkov radiation. Due to the numerical instability the nuclei are not able to propagate stably unless a very small lattice spacing is chosen along $z$, which reduces the mismatch in propagation velocities. Fortunately, these numerical problems can be solved by modifying the lattice discretization of the color fields. In \cite{Ipp:2018hai} we show how an improved numerical scheme restores the correct propagation of wave modes along the beam axis such that artificial Cherenkov radiation is effectively avoided.
This modification mainly amounts to replacing specific spatial plaquette terms (see fig.~\ref{fig:wilson} (a)) with time-averaged generalizations of these terms. Figure \ref{fig:wilson} (b) shows one example of such a time-averaged term.
In comparison to the leapfrog scheme eqs.~\eqref{eq:lattice_1} and \eqref{eq:lattice_2}, which is an explicit finite difference method, our numerical method is in the form of a system of semi-implicit eqs., which is solved in an iterative manner. Our semi-implicit method therefore trades computational performance for numerical stability and accuracy.

\section{Results}
Here we review the most important results that have been obtained from simulations of collisions of nuclei with finite longitudinal extent using the standard leapfrog scheme \cite{Gelfand:2016yho,Ipp:2017lho,Ipp:2017uxo}.
In these simulations we use initial conditions of the form eq.~\eqref{eq:single_sheet}, where $\lambda$ is a Gaussian of width $L$ along $z$. We also assume $\mu^2$ to be constant in the transverse plane. The longitudinal thickness $L$ has been set to $L=m_{N}R/\sqrt{s_{NN}}$ with collision energy $\sqrt{s_{NN}}$, nuclear radius $R$ and nucleon mass $m_{N}\approx1\:\mbox{GeV}$.
The saturation momentum $Q_s$ grows with collision energy as $Q_{s}^{2}\approx\left(\sqrt{s_{NN}}\right)^{0.25}\,\mbox{GeV}^{2}$ \cite{Lappi:2006hq,Lappi:2007ku,Kharzeev:2001gp}.
The MV model parameter $\mu$ can be determined from $0.75\, g^{2}\mu\simeq Q_{s}$ with coupling $g\approx2$ \cite{Schenke:2012hg}.
The ultraviolet modes are regulated by $\Lambda=10\:\mbox{GeV}$, and we vary the infrared regulator $m$ in the range from 0.2 to $0.8\,\mbox{GeV}$ to check for its dependency. The simulations presented use the gauge group $\mbox{SU}(2)$ instead of SU$(3)$, which should give qualitatively comparable results \cite{Ipp:2010uy}.
The simulations have been performed on a lattice with $2048 \times 192^2$ cells with finer resolution along the longitudinal direction. The simulation box corresponds to a volume of $\left(6\,\mbox{fm}\right)^{3}$.

\begin{figure}
\resizebox{0.48\textwidth}{!}{
\centering
\includegraphics{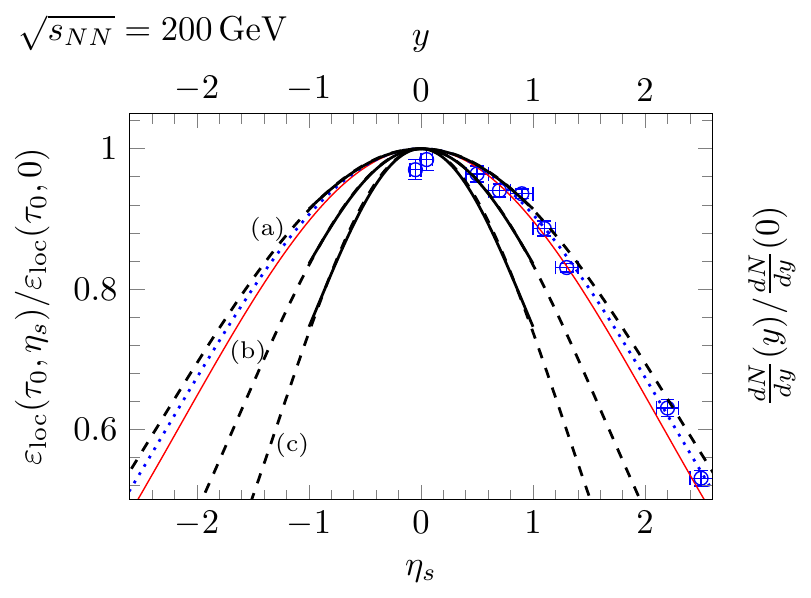}
}
\caption{Comparison between simulation and experimental results for the space-time rapidity profile \cite{Ipp:2017lho}.
The thick black solid lines show simulation results of the local rest frame energy density $\e_{\text{loc}}(\tau_{0},\eta_{s})$ at $\tau_{0}=1\,\mbox{fm}$/c for various values of the infrared regulator $m$.
The following widths of the Gaussian profiles have been extracted: 
(a) $m=0.2\,\mbox{GeV}$ with $\sigma_{\eta}=2.34$, (b) $m=0.4\,\mbox{GeV}$ with $\sigma_{\eta}=1.66$ and (c) $m=0.8\,\mbox{GeV}$ with $\sigma_{\text{\ensuremath{\eta}}}=1.28$.
For comparison, the experimentally obtained profile of $\pi^{+}$ multiplicity $dN/dy$ at RHIC is given by the blue data points \cite{Bearden:2004yx}, with a width of $\sigma_{\text{exp}}=2.25$.
The thin red line corresponds to the profile predicted by the Landau model with $\sigma_{\text{Landau}}=\sqrt{\ln\gamma}\approx2.15$.
\label{fig:RHIC200}}
\end{figure}

The main observables can be obtained from the components of the energy-momentum tensor $T^{\mu\nu}$ which are extracted from the gluon fields of the simulation. We average over $15$ collision events to obtain the expectation value $\ev{T^{\mu\nu}}$.
Due to the symmetries of the MV model, certain components vanish, and the remaining contributions of the averaged energy-momentum tensor are given by
\begin{equation}
\ev{T^{\mu\nu}}=\left(\begin{array}{cccc}
\ev{\e} & 0 & 0 & \ev{S_{L}}\\
0 & \ev{p_{T}} & 0 & 0\\
0 & 0 & \ev{p_{T}} & 0\\
\ev{S_{L}} & 0 & 0 & \ev{p_{L}}
\end{array}\right),
\end{equation}
where $\ev{\varepsilon}$ is the energy density in the laboratory frame, $\ev{p_{L}}$ and $\ev{p_{T}}$ are the longitudinal and transverse pressure components and $\ev{S_{L}}$ is the longitudinal component of the Poynting vector.
The energy density $\ev{\varepsilon}$ as given in the laboratory frame is depicted in fig.\,\ref{fig:fig1}.
By diagonalizing the energy-momentum tensor, we obtain the local rest frame energy density $\ev{\e_{\text{loc}}}$, which can be expressed in terms of proper time $\tau=\sqrt{t^{2}-z^{2}}$ and space-time rapidity $\eta_{s}=\ln\left[\left(t-z\right)/\left(t+z\right)\right] / 2$.

Figure \ref{fig:RHIC200} shows the local energy density $\varepsilon_{loc}(\tau_0, \eta_s)$ as a function of space-time rapidity $\eta_{s}$ for $\sqrt{s_{NN}}=200\,\mbox{GeV}$.
The black solid lines show the rapidity profiles as extracted from of our simulations, which can be fitted to a Gaussian shape (dashed continuing lines). 
The profiles have been extracted at $\tau_{0}=1\,\mbox{fm}/c$ where the Glasma turns into the QGP.
Already at times $\tau_{0}\gtrsim0.3\,\mbox{fm}/c$, the system enters a free-streaming evolution with longitudinal velocity $v_z \approx z / t$ and the shape of the profile does not change anymore \cite{Ipp:2017uxo}. 
The width of the profiles depends on the energy $\sqrt{s_{NN}}$ and becomes flatter with increasing energy as expected from the recovery of boost invariance at higher energies \cite{Ipp:2017lho}.
We also find a strong dependency on the infrared regulator $m$, where higher values of $m$ make the rapidity profiles narrower. 
While this strong dependence on the infrared regulator seems unexpected, it may indicate that the screening length $\lambda_D \propto 1/m$ plays an important role in generating a deviation from boost invariance. It is not only the longitudinal thickness $L$, but the dimensionless ratio $L / \lambda_D$, which seems to determine the shape of the profiles.

Interestingly, the simulation results agree well with measured rapidity profiles of pion multiplicities at RHIC \cite{Bearden:2004yx} which are indicated by the blue data points in figure \ref{fig:RHIC200}. At these energies, the experimental results also agree with the Gaussian rapidity profile as obtained by the hydrodynamic Landau model \cite{Landau:1953gs} where the width is given by $\sigma_{\text{Landau}}=\sqrt{\ln\gamma}$ with the Lorentz gamma factor $\gamma$.
For other particle species, the experimental momentum rapidity distribution of particle multiplicities deviates from the Landau picture but can still be fitted to a Gaussian distribution with slightly larger width \cite{Abbas:2013bpa}.
However, it should be noted that especially at higher energies the Landau model predicts rapidity profiles which are too narrow as measured by the \mbox{ALICE} collaboration \cite{Abbas:2013bpa}. Whether our simulations of the 3+1D Glasma can describe these wider profiles at LHC energies will be the topic of future work.

It is important to note that figure \ref{fig:RHIC200} shows the rapidity profiles of two different quantities: the local energy density $\varepsilon_\mathrm{loc}(\tau, \eta_s)$ as a function of space-time rapidity $\eta_s$ and the distribution of particle multiplicities $dN/dy$ as a function of momentum rapidity $y$.  This comparison is justified because our simulations show \cite{Ipp:2017lho, Ipp:2017uxo} that the 3+1D Glasma settles into a state of free-streaming flow $v_z \approx z / t$ and vanishing longitudinal pressure $p_L \approx 0,$ similar to the 2+1D Glasma. Free-streaming flow implies that -- ignoring a subsequent hydrodynamical phase -- we can identify momentum rapidity $y$ with space-time rapidity $\eta_s$, similar to the case of the Bjorken model \cite{Bjorken:1982qr}. It also implies that $T^{\mu\nu}$ becomes diagonal in the $(\tau, \eta_s)$ frame. Due to vanishing longitudinal pressure, the energy-momentum tensor in the rest frame is anisotropic and reads
\begin{equation}
\ev{T^{\mu\nu}} = \mathrm{diag}(\ev{\varepsilon_\mathrm{loc}}, \, \ev{\varepsilon_\mathrm{loc}} / 2, \, \ev{\varepsilon_\mathrm{loc}} / 2, \, 0)^{\mu\nu},
\end{equation}
where we used $2 \, p_T = \varepsilon_\mathrm{loc}$ due to conformal symmetry. In contrast to the Bjorken model, there are \textit{a priori} no restrictions on the rapidity dependence of the energy density $\varepsilon_\mathrm{loc}(\tau, \eta_s)$. This can be checked using  energy-momentum conservation
\begin{equation} \label{eq:em_conservation}
\nabla_\mu T^{\mu\nu} = 0,
\end{equation}
which simply reduces to
\begin{equation} \label{eq:late_time_energy}
\partial_\tau \varepsilon_\mathrm{loc}(\tau, \eta_s) = - \varepsilon_\mathrm{loc}(\tau, \eta_s) / \tau.
\end{equation}
Equation \eqref{eq:late_time_energy} leads to $\varepsilon_\mathrm{loc} \propto 1 / \tau$, but does not impose any additional constraints on the rapidity dependence of the energy density. On the other hand, the Bjorken model requires that $T^{\mu\nu}$ is isotropic in the rest frame, which combined with eq.~\eqref{eq:em_conservation}, then leads to boost invariance. As a first approximation, it is therefore reasonable to compare the space-time rapidity profile of $\varepsilon_\mathrm{loc}$ to the momentum rapidity profile of charged particles as shown in figure \ref{fig:RHIC200}. For a more quantitative comparison, our results should be used as input for a subsequent hydrodynamic simulation, which may slightly increase the width of the profiles \cite{Schenke:2016ksl}.

\begin{figure}
\resizebox{0.48\textwidth}{!}{
\centering
\includegraphics{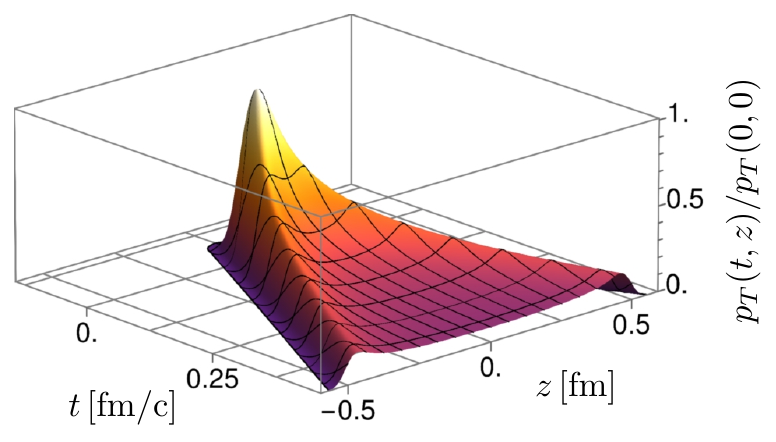}
}
\caption{Space-time distribution of the normalized transverse pressure $\ev{p_{T}(t,z)}/\ev{p_{T}(0,0)}$ \cite{Ipp:2017lho}. 
The parameters are the same as in figure \ref{fig:RHIC200} with $m=0.2\,\mbox{GeV}$.
The transverse pressure corresponds to longitudinal chromo-magnetic and -electric fields and thus to the the longitudinal component of the energy density $\ev{\e_{L}(t,z)}$.
Contrary to the boost-invariant case, this quantity falls off steeply along the boundary of the light cone.
\label{fig:ep_L_spacetime}}
\end{figure}

To better understand how rapidity dependence of observables develops in our simulations, we can look at the transverse pressure $\ev{p_{T}(z,t)}$ in figure \ref{fig:ep_L_spacetime}.
From the energy-momentum tensor, one finds that the transverse pressure is linked to the longitudinal fields of the Glasma, whereas longitudinal pressure involves both, transverse and longitudinal field components \cite{Gelfand:2016yho}
\begin{eqnarray}
\ev{p_{T}} & = & \frac{1}{2}\ev{E_{L}^{2}+B_{L}^{2}},\label{eq:pT_eL}\\
\ev{p_{L}} & = & \frac{1}{2}\ev{E_{T}^{2}+B_{T}^{2}-E_{L}^{2}-B_{L}^{2}},
\end{eqnarray}
where the square implies a summation over color indices.
In the boost invariant case, the initial conditions of the Glasma are specified at $\tau=0$. This corresponds to the boundary of the forward light cone, where the longitudinal fields would be constant.
In contrast, in our simulation the longitudinal fields are peaked around the collision region at $t\sim z\sim0$ and decrease rather quickly along the light cone boundaries. 
Accordingly, there is less Glasma being produced at larger values of rapidity $\eta_{s}$ which produces the Gaussian profiles.
It is interesting to see that our weak coupling results are in qualitative agreement with strong coupling results from holographic models of heavy-ion collisions that also exhibit similar transverse pressure distributions  \cite{Casalderrey-Solana:2013aba,vanderSchee:2015rta}.

\section{Conclusions and outlook}

In this paper we reviewed our progress on 3+1D Glasma simulations.
Our simulations allow to explore the creation of the Glasma in heavy-ion collisions beyond the commonly assumed boost-invariant case.
We do this by introducing a finite longitudinal extent for the incoming nuclei corresponding to realistic Lorentz contractions as found for example at RHIC.
Without the usual simplifications of boost-invariance, we have to keep the color currents of the hard partons in the Glasma simulation.
This is achieved using CPIC in the laboratory frame.
Using the MV model, we demonstrated that our approach can give rise to Gaussian rapidity profiles in the energy density. 
These profiles depend on the energy of the incoming nuclei, but also on an infrared regulator. 
For energies used at RHIC we obtain qualitative agreement of these profiles \cite{Ipp:2017uxo}.
This is remarkable as it shows that boost invariance can be broken already at the classical level if the longitudinal structure is properly taken into account.
This nicely complements findings from holographic models where similar profiles can be found \cite{Casalderrey-Solana:2013aba,vanderSchee:2015rta}.

Algorithmic improvements in the form of a new semi-implicit solver \cite{Ipp:2018hai} will allow for further explorations of our boost-invariance breaking simulations.
By modifying the standard Wilson gauge action we achieve a dispersion-free propagation along the longitudinal direction which cures the numerical Cherenkov instability which has plagued previous simulations.
This sets the basis for more accurate and larger simulations valid for larger ranges of rapidity which are necessary for a comparison of rapidity profiles at LHC collision energies.
One crucial aspect that shall be studied in this context is the role of longitudinal color fluctuations.
These are usually approximated as an infinitely thin stack of uncorrelated sheets of color charge \cite{Fukushima:2007ki}.
Dispersion-free propagation will allow for the fine-grained simulation of collisions and the study of the effect of internal longitudinal color structures on the creation of the Glasma.
In principle, it should be straightforward to also include more realistic sub-nucleonic color structure in the transverse direction as is the case in the IP-Glasma model \cite{Schenke:2012wb,Schenke:2012hg}.
Here, one is essentially limited by the large computational requirements of such three-dimensional simulations.

On a more conceptual level, it would be interesting to better understand the relation between the boost-invariance breaking that we find at the leading classical order and a similar breaking that can be found at next-to-leading order from the JIMWLK evolution  \cite{Schenke:2016ksl}.
It would be a highly desirable but presumably very non-trivial task to generalize the JIMWLK evolution eqs.~to be applicable to three-dimensional color distributions.
Another extension to our work would be the deviation from the eikonal approximation and the inclusion of dynamical colored particles. This could also be a way to accommodate three-dimensional extensions of the calculation of quantities like energy loss or momentum broadening from the Glasma \cite{Ipp:2020mjc}.

\section{Acknowledgement}
The authors thank for a Short Term Scientific Mission (STSM) to CEA Saclay
for scientific discussions with Jean-Paul Blaizot, Fran\c{c}ois Gelis and Edmond Iancu
within the framework of the COST action CA15213 ``Theory of hot matter and relativistic heavy-ion collisions (THOR)''.
This work has been supported by the Austrian Science Fund FWF No. P28352. 

\bibliographystyle{spphys.bst}
\bibliography{references.bib}

\end{document}